\pdfoutput=1
\documentclass[a4paper]{article}

\usepackage{amsmath}
\usepackage{caption}
\usepackage{latexsym}
\usepackage[pdftex]{graphicx}

\usepackage{color}
\usepackage{soul}
\usepackage{secdot}
\sectiondot{section}
\sethlcolor{white}

\usepackage[left,columnwise]{lineno}

\usepackage{times}
\usepackage{bm}
\usepackage{natbib}

\usepackage[plain,noend]{algorithm2e}
\usepackage{listings}
\newcommand*\patchAmsMathEnvironmentForLineno[1]{%
  \expandafter\let\csname old#1\expandafter\endcsname\csname #1\endcsname
  \expandafter\let\csname oldend#1\expandafter\endcsname\csname end#1\endcsname
  \renewenvironment{#1}%
     {\linenomath\csname old#1\endcsname}%
     {\csname oldend#1\endcsname\endlinenomath}}%
\newcommand*\patchBothAmsMathEnvironmentsForLineno[1]{%
  \patchAmsMathEnvironmentForLineno{#1}%
  \patchAmsMathEnvironmentForLineno{#1*}}%
\AtBeginDocument{%
\patchBothAmsMathEnvironmentsForLineno{equation}%
\patchBothAmsMathEnvironmentsForLineno{align}%
\patchBothAmsMathEnvironmentsForLineno{flalign}%
\patchBothAmsMathEnvironmentsForLineno{alignat}%
\patchBothAmsMathEnvironmentsForLineno{gather}%
\patchBothAmsMathEnvironmentsForLineno{multline}%
}

\begin{document}

\large
\begin{center}
{\LARGE Effect sizes of the differences between means without assuming the variance equality and between a mean and a constant}
\newline

Satoshi Aoki$^{1,*}$
Motomi Ito$^{2}$, and
Masakazu Shimada$^{2}$\\
\end{center}

$^{1}$Department of Biological Sciences, the University of Tokyo.

$^{2}$Department of General Systems Studies, the University of Tokyo.

*email: aoki171@g.ecc.u-tokyo.ac.jp

\begin{center}
\section*{Abstract}
\end{center}
Hedges' $d$, an existing unbiased effect size of the difference between means, assumes the variance equality. However, the assumption of the variance equality is fragile, and is often violated in practical applications. Here, we define $e$, a new effect size of the difference between means, which does not assume the variance equality. In addition, another novel statistic $c$ is defined as an effect size of the difference between a mean and a known constant. Hedges' $g$, our $c$, and $e$ correspond to Student's unpaired two-sample $t$ test, Student's one-sample $t$ test, and Welch's $t$ test, respectively. An R package is also provided to compute these effect sizes with their variance and confidence interval.\newline

Keywords: Cohen's $d$; Confidence interval; Constant; Effect size; Hedges' $d$; Hedges' $g$.

\newpage

\setcounter{section}{0}
\section{Introduction}
\label{s:intro}
\begin{newblock}\sethlcolor{cyan}\hl{}\end{newblock}
An effect size is a term which refers to various kinds of parameters or statistics to define or measure the magnitude of effects. This study treats effect sizes of the difference which treat the magnitude between two means or between a mean and a constant. In general, effect sizes are used to estimate the magnitude of effect independent of the sample size (Nakagawa and Cuthill, 2007), to compare the results of multiple studies (meta-analysis; (Glass, 1976)), or to determine the statistical power or the appropriate sample size (power-analysis; (Cohen, 1988)). In spite of such importance of effect sizes, the existing effect sizes of the difference assume the equality of the variance which is practically hard to assume. In addition, an effect size of the difference between a mean and a constant was found to be undefined. To solve these problems, we defined an effect size of the difference between means which does not assume the variance equality based on Welch's $t$ test (Welch, 1938, 1947). Also, we defined an effect size of the difference between a mean and a constant based on one sample $t$ test (Fisher, 1925). 

The rest of this paper is organized as follows. In section 2, the existing effect sizes of the difference are introduced. In section 3, we define two new effect sizes, their variance, and confidence interval. In section 4, we introduce our new R package to compute the effect sizes, using some examples. In section 5, we discuss the nature and application range of the new effect sizes.

\section{Existing effect sizes of the difference}
\label{s:existings}
Glass (1976) was the first person that suggested an effect size of the difference. He defined it as ``the mean difference on the outcome variable between treated and untreated subjects divided by the within group standard deviation.'' He clearly distinguished the treated (experimental) group from the untreated (control) group, and there was no assumption about the two groups. His effect size was subsequently formulated and named Glass' $\Delta$ by Hedges (1981), which is
\begin{equation}
\label{GlassDelta}
\Delta = (\bar{Y}^E - \bar{Y}^C)/S^C, 
\end{equation}
where $\bar{Y^E}$ is the mean of the variable in the experimental group, $\bar{Y^C}$ is that in the control group, and $S^C$ is the unbiased standard deviation of the control group.

After this invention of an effect size of the difference, Hedges (1981) defined the other effect size as a parameter for independently and normally distributed two populations, $Y^1\sim N(\mu_1,\sigma^2)$ and $Y^2\sim N(\mu_2,\sigma^2)$:
\begin{equation}
\label{delta}
\delta = (\mu_1-\mu_2)/\sigma.
\end{equation}
Note that both populations share the common variance $\sigma^2$. As the estimator of this parameter, the other effect size of the difference is represented as $g$ by Hedges (1981) and $d$ by Cohen (1988). These two effect sizes are equivalent, and they are defined for the equally handled two groups with equal variances.  Although this effect size was defined by Hedges (1981) earlier, this effect size is usually called Cohen's $d$. However, we chose to call it $g$ in this article in order to distinguish it from the other $d$ which we will introduce later. The statistic $g$ is defined as
\begin{equation}
\label{one}
g = (\bar{Y}^1 - \bar{Y}^2)/S^\text{pooled},
\end{equation}
where
\begin{equation*}
S^\text{pooled} = \sqrt{\frac{s_1^2(n_1-1) + s_2^2(n_2-2)}{n_1 + n_2 - 2}},
\end{equation*}
and
\begin{equation}
\label{svar}
s_i^2 = \frac{\sum_{j=1}^{n_i} (Y^i_j - \bar{Y^i})^2}{n_i-1}.
\end{equation}
Here, $\bar{Y}^1$, $Y^1_j$, and $n_1$ are the mean of the sample, the sample (random variable), and the sample size of group 1, while $\bar{Y}^2$, $Y^2_j$, and $n_2$ are those of group 2. For the denominator, this effect size uses the pooled standard deviation, which suggests the most precise population variance under the assumption of the equal variance (Hedges, 1981).

The $g$ (3) is biased from $\delta$ (2), making it unsuitable for analyses that do not treat the whole population. The unbiased estimator of $\delta$ (2) is defined as $g^U$ in Hedges (1981) and $d$ in Hedges and Olkin (1985). In this article, we call it $d$, which is
\begin{equation}
\label{two}
d = J(n_1 + n_2 - 2)g.
\end{equation}
By using the gamma function, the correction coefficient $J$ is defined as
\begin{equation}
\label{three}
J(m) =\frac{\Gamma(m/2)}{\sqrt{m/2}\Gamma\{(m - 1)/2\}}.
\end{equation}
The effect sizes $g$ (3) and $d$ (5) are widely used in various regions of sciences, but they assume the variance equality just like Student's t test (Student, 1908; Fisher, 1925). In the next section, we propose a new effect size of the difference which does not assume the variance equality as well as another new effect size of the difference between a mean and a constant.

\section{Proposed effect sizes}
\label{s:dif}

\subsection{An effect size of the difference between means without assuming the variance equality}
\label{s:e}
First, we define the parameter of an effect size of the difference between means for two independently and normally distributed populations $N_1(\mu_1,\sigma_1^2 )$ and $N_2(\mu_2,\sigma_2^2)$ as
\begin{equation}
\label{epsilon}
\epsilon_r=\frac{\mu_1-\mu_2}{\sqrt{(\sigma_1^2+r\sigma_2^2)/(r+1)}},
\end{equation}
where $r$ is a non-negative real number. This parameter is not generalization of $\delta$ (2), and is different from it. Then, suppose two independently and normally distributed populations with the samples $Y^1_i$ ($i= 1,...,n_1$) and $Y^2_i$ ($i= 1,...,n_2$), and the sample mean $\bar{Y}^1$ and $\bar{Y}^2$. Based on the statistic $t_w$ so-called Welch's $t$ (Welch,1938, 1947), an biased estimator of $\epsilon_r$ (7) is defined as
\begin{equation}
\label{four}
e^\text{biased}=t_w/\sqrt{\tilde{n}},
\end{equation}
where
\begin{equation}
\label{welcht}	
t_w=\frac{\bar{Y}^1 - \bar{Y}^2}{\sqrt{s_1^2/n_1 + s_2^2/n_2}},
\end{equation}
$s_i^2$ is the same as (4), and
\begin{equation}
\label{enye}
\tilde{n}=n_1n_2/(n_1+n_2).
\end{equation}
Finally, $e$, the unbiased estimator of $\epsilon_r$ (7), is
\begin{equation}
\label{five}
e=e^\text{biased}J(f).
\end{equation}
Therefore,
\begin{equation*}
{\rm E}(e)=\epsilon_r.
\end{equation*}
Here, $r$ corresponds to the ratio $n_1/n_2$. $J$ is the correction coefficient which is defined in equation (6). The degree of freedom $f$ is approximately calculated by using the Welch-Satterthwaite  equation (Welch, 1938; Satterthwaite, 1941) as
\begin{equation}
\label{six}
f = \frac{(s_1^2/n_1+s_2^2/n_2)^2}{s_1^4/\{n_1^2(n_1-1)\}+s_2^4/\{n_2^2(n_2-1)\}}.
\end{equation}
The variance of $e$ (11) is
\begin{equation*}
{\rm var}(e)=\frac{f}{f-2}J^2(f)\{1/\tilde{n}+\epsilon_r^2\}-\epsilon_r^2.
\end{equation*}
Although this effect size is derived from the difference, we dare to name it not $d$ but $e$. This is because Cohen's $d$ (3) and Hedges' $d$ (5) already exist, and more $d$ would cause more confusion. The proof of the bias correction and variance derivation does not assume the variance equality (see the Appendix). In addition, $e$ (11) is a consistent estimator of $\epsilon_r$ (7) at the same time. See the Appendix for the proof of the consistency. 

\subsection{An effect size of the difference between a mean and a known constant}
As a parameter, an effect size of the difference between a mean and a constant is defined for a normally distributed population $N_1(\mu, \sigma_1^2 )$ and a known constant $C$ as
\begin{equation}
\label{gamma}
\gamma=(\mu_1-C)/\sigma_1.
\end{equation}
Next, an effect size as the statistic is defined for a normally distributed population with the sample value $Y^1_i$ ($i = 1,$ ... ,$n_1$), the sample mean $\bar{Y^1}$, and a known constant $C$ as
\begin{equation}
\label{seven}
c^\text{biased}=(\bar{Y}^1 - C)/s_1.
\end{equation}
The $s_1$ is the square root of (4). Then, using $c^\text{biased}$(14), the unbiased estimator of the effect size parameter $\gamma$ (13) is
\begin{equation}
\label{eight}
c=c^\text{biased}J(n_1-1).
\end{equation}
Therefore,
\begin{equation*}
{\rm E}(c)=\gamma.
\end{equation*}
The correction coefficient $J$ (6) is the same one as used above. The variance of $c$ is
\begin{equation*}
{\rm var}(c)=\frac{n_1-1}{n_1-3}J^2(n_1-1)(\frac{1}{n_1-1}+\gamma^2)-\gamma^2.
\end{equation*}
See the Appendix for proofs of the bias correction and the derivation of the variance. In addition, $c$ (15) is a consistent estimator of $\gamma$ (13) (See the Appendix for the proof). When interested in constants rather than variables, you can use $c'$ defined as
\begin{equation*}
c'=(C-\bar{Y}^1)J(n_1-1)/s_1
\end{equation*}
instead of $c$. What is important is not to confuse $c$ (15) and $c'$.

\subsection{Confidence intervals of effect sizes}
The confidence interval (CI) of effect sizes of the difference is not directly given by a formula (Cumming and Finch, 2001). The CI is derived from that of noncentral parameters of noncentral t-distribution, which is in turn gained by some searching method. The CI of the biased effect sizes are given as:
\begin{equation*}
[ncp_L/\sqrt{\tilde{n}},\:\:ncp_H/\sqrt{\tilde{n}}]\:\:for\:\:g,
\end{equation*}

\begin{equation*}
[ncp_L/\sqrt{\tilde{n}},\:\:ncp_H/\sqrt{\tilde{n}}]\:\:for\:\:e^\text{biased},
\end{equation*}
and
\begin{equation*}
[ncp_L/\sqrt{n_1-1},\:\:ncp_H/\sqrt{n_1-1}]\:\:for\:\:c^\text{biased},
\end{equation*}
where $ncp_L$ is the noncentral parameter which gives the upper limit of cumulative probability (e.g. 0.975 cumulative probability for 95 \% CI) for noncentral t-distribution with the corresponding t value (see section 5) and the degree of freedom, and $ncp_H$ is that which gives the lower limit (e.g. 0.025 cumulative probability for 95 \% CI), and $\tilde{n}$ and $n_1$ are the same as (10) and (14). The CIs for the unbiased estimator of the effect sizes are given by multiplying the corresponding correction coefficient $J$ (6) of the corresponding degree  of freedom to the above intervals. We do not discuss the practical usage of these CIs in this article, because it has already been discussed in the other studies, such as Nakagawa and Cuthill (2007).   

\subsection{Practical application}
While the situation to use $c$ (15) is clearly different from that to use $d$ (5), the $e$ (11) and $d$ (5) have a similar application range in practice. Therefore, we prepared an example of the applications of $e$ (11) compared to $d$ (5). Table 1 shows famous data of three \textit{Iris} species by Fisher (1936), which has various variances. Note that only the petal width of \textit{I. setosa} has fewer significant digits. For this data, we calculated $d$ (5), $e$ (11), the ratio of $d$ (5) to $e$ (11), and the ratio of the standard deviations of the two comparing data. Theoretically, $e$ (5) is more precise estimator of its own parameter than $d$ (11) in this non-equal variance situation. 
\begin{table}[hpt]
\vspace*{0cm}
\caption{Measured characteristics of three \textit{Iris} species shown in Fisher (1936). The last two raws show the average and the standard deviation of the corresponding column. The numbers show the lengths in centimeter. S.L.: Sepal length. S.W.: Sepal width. P.L.: Petal length. P.W.: Petal width.}
\vspace*{-0.3cm}
\includegraphics[width=9.5cm,pagebox=cropbox,clip]{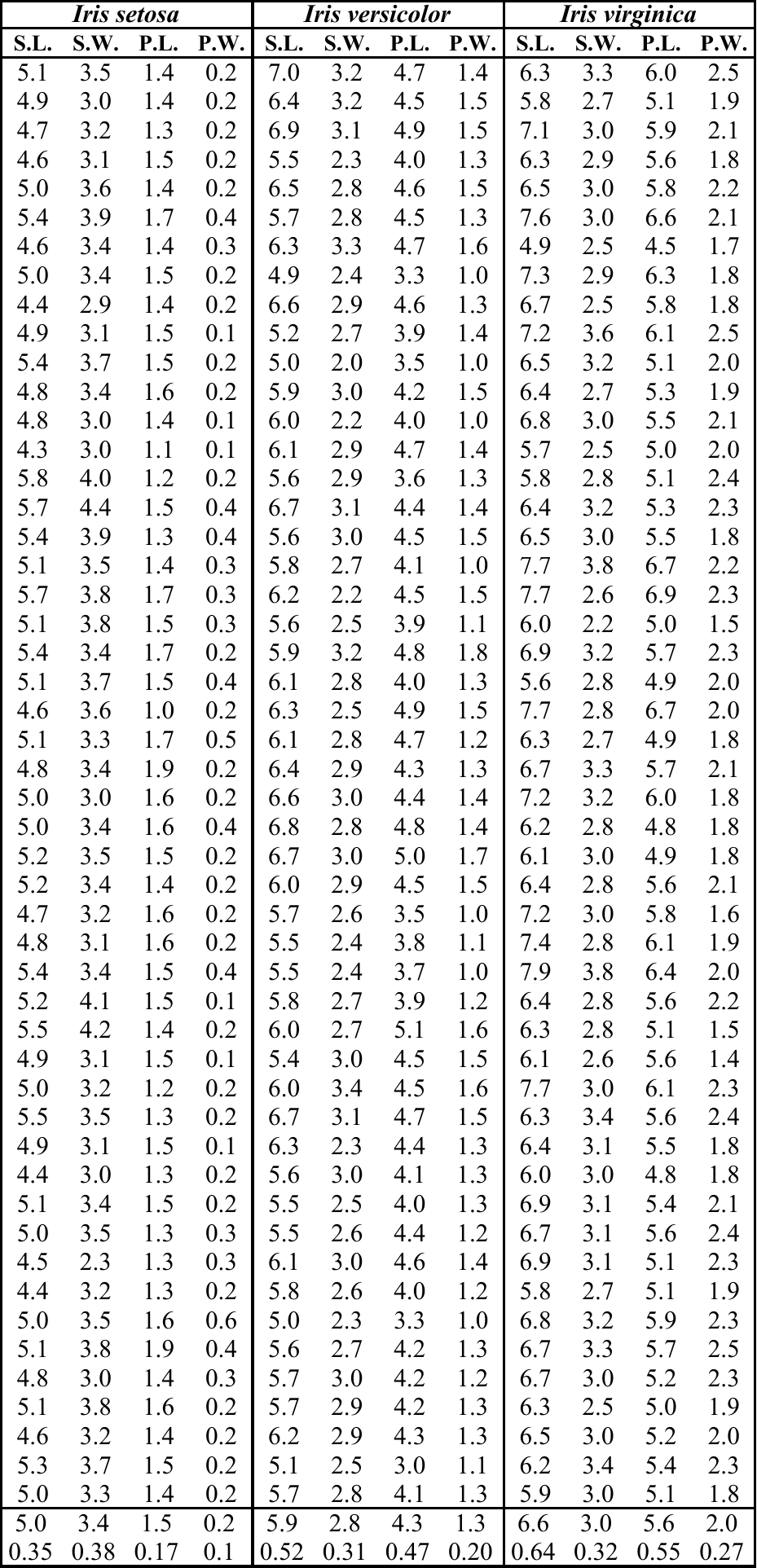}

\label{DataTable1}
\end{table}

The calculated result is shown in Table 2. When considering their significant digits, the comparing pair of the sepal length of \textit{I. setosa} and \textit{I. virginica} showed the different effect size of $d$ (5) and $e$ (11). (Bolds in Table 2.) Even though most pairs showed identical values of $d$ (5) and $e$ (11), the result showed that these two effect sizes can be different even in two significant digits. 

\begin{table}[hpt]
\vspace*{0cm}
\caption{Calculated effect sizes of the difference for the data shown in Table 1.
Chara.: Characteristics. S.L.: Sepal length. S.W.: Sepal width. P.L.: Petal length. P.W.: Petal width. Taxa: Compared taxa. 1: \textit{I. setosa}. 2: \textit{I. versicolor}. 3: \textit{I. virginica}. d: Effect size $d$ (5). e: Effect size $e$ (11). These effect sizes are shown in the original significant digits. d/e: The ratio of $d$ (5) to $e$ (11) calculated without considering the significant digit. sd ratio: The ratio of the standard deviations of the compared data calculated without considering the significant digit. Note that the reverse comparisons, such as 2 vs 1, were also conducted, but omitted from this table. This is because their effect sizes are the opposites of the original values, and d/e and sd ratio are the inverses of the original ones.}
\vspace*{0cm}
\includegraphics[width=10cm,pagebox=cropbox,clip]{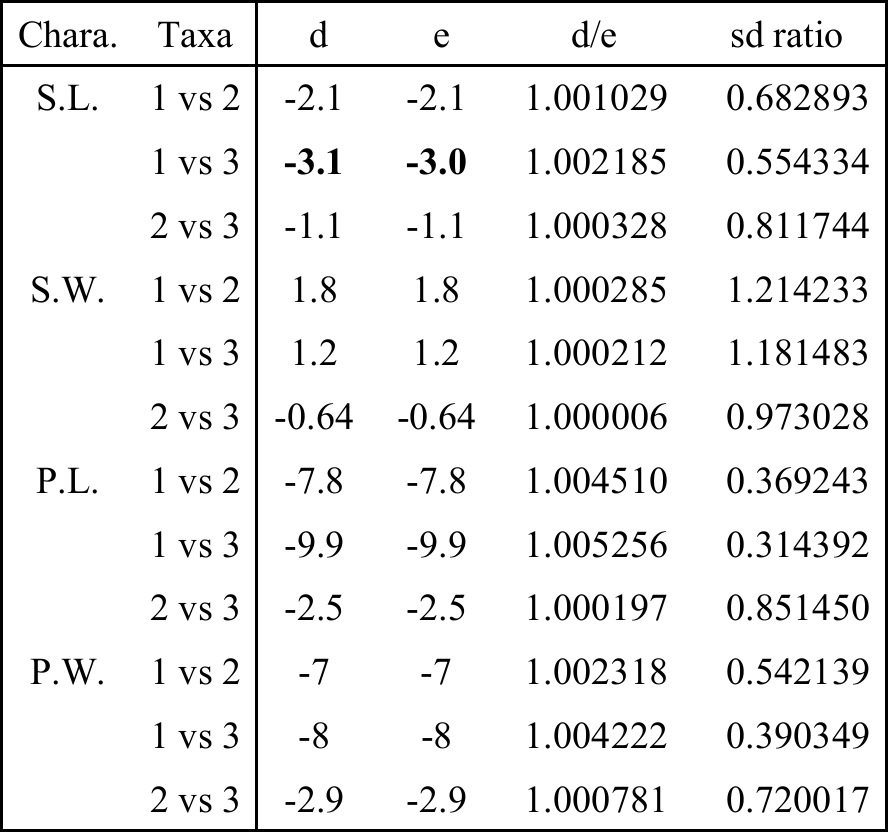}

\label{CalculatedTable1}
\end{table}

Figure 1 shows the ratio of $d$ (5) to $e$ (11) plotted against the ratio of standard deviations of the comparing data. This figure shows that the similar two standard deviations give similar $d$ (5) and $e$ (11). In other words, the more different two  standard deviations more encourage the use of $e$ (11).
\begin{figure}[hpt]
\begin{flushleft}
\includegraphics[width=10cm,pagebox=cropbox,clip]{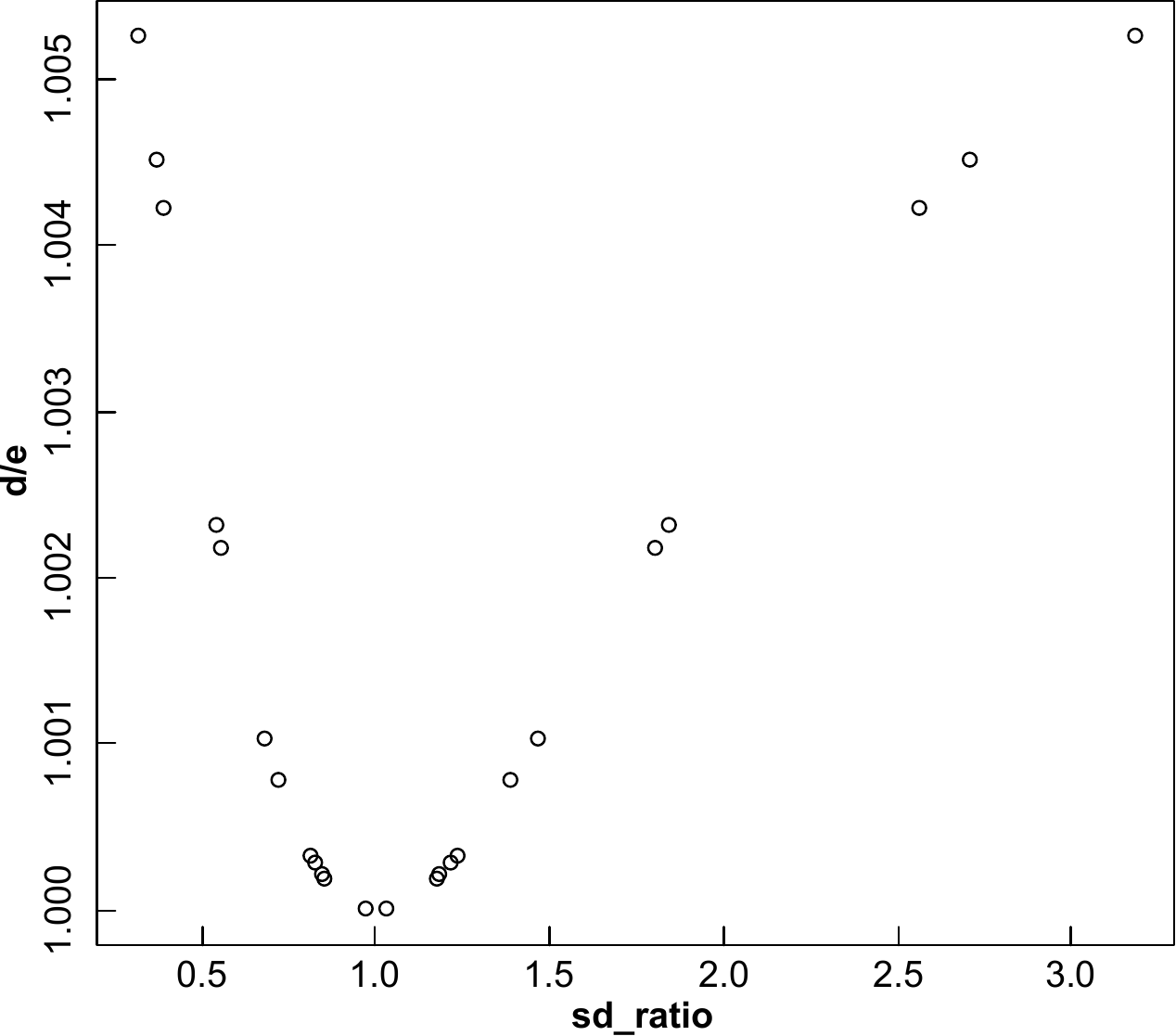}
\end{flushleft}
\caption{Plotted graph of Table 2. d/e: The ratio of $d$ (5) to $e$ (11).  sd\_ratio: The ratio of the standard deviations of the compared data.}
\label{DdivE}
\end{figure}

\section{New R package \& its application}
\label{app}
A new package `es.dif' for R (R core team, 2019) is provided. It enables to compute the statistics $d$ (5),  $e$ (11), $c$ (15), their biased statistics, variance, and CI based on the two samples or the statistics (mean, variance, and sample size) of the samples. In this package, approximation of $J$ (6) (Hedges, 1981) is not employed unless its degree of freedom exceeds 342, when the gamma function returns too large values to be treated in R. The CI is gained by binary search. The figure on this article is drawn with this package.

This section hereafter shows some examples of the package. First, the following script calculates d (5), e (11), their variances and 95\% CIs for data 1 (0,1,2,3,4) and data 2 (0,0,1,2,2).

\noindent
\texttt{\normalsize
> library(es.dif)
\newline
> data1<-c(0,1,2,3,4)
\newline
> data2<-c(0,0,1,2,2)
\newline
> es.d(data1,data2)
\newline
     [,1]         [,2]
\newline
[1,] "Hedges' d:" "0.682379579593354"
\newline
[2,] "variance:"  "0.484026380702367"
\newline
[3,] "CI:"        "[ -0.503527216375147 , 1.82938058482178 ]"
\newline
> es.e(data1,data2)
\newline
     [,1]          [,2]
\newline
[1,] "Unbiased e:" "0.668264936033828"
\newline
[2,] "variance:"   "0.506830833214916"
\newline
[3,] "CI:"         "[ -0.50334965496395 , 1.7965317007171 ]"
}

Using options of the function, you can change the type I error rate for the CI, calculate biased effect sizes, and output results in the vector style. For example, $c^\text{biased}$ (14) with 99\% CI in the vector style is calculated by this script.

\noindent
\texttt{\normalsize
> library(es.dif)
\newline
> data1<-c(0,0,1,2,2)
\newline
> data2<-c(2)
\newline
> es.c(data1,data2,alpha=0.01,unbiased=FALSE,vector\_out=TRUE)
\newline
[1] -1.0000000  0.9292037 -2.5390625  0.5778885
}

In the vector-style output, the four values in the vector show the effect size, its variance, lower limit of the CI, and higher limit of the CI. In addition, this package includes the functions which can output effect sizes from the (estimated) parameters and the sample sizes. Following scripts compute $d$ (5) and $e$ (11) for two populations, $N(1, 2)$ and $N(0, 1)$ with the sample size 5 and 10.

\noindent
\texttt{\normalsize
> library(es.dif)
\newline
> mean1<-1
\newline
> mean2<-0
\newline
> var1<-2
\newline
> var2<-1
\newline
> n1<-5
\newline
> n2<-10
\newline
> es.para.d(mean1,mean2,var1,var2,n1,n2)
\newline
     [,1]         [,2]                                       
\newline
[1,] "Hedges' d:" "0.82286529714397"                         
\newline
[2,] "variance:"  "0.349443397657368"                        
\newline
[3,] "CI:"        "[ -0.248827687382689 , 1.86616833367494 ]"
\newline
> es.para.e(mean1,mean2,var1,var2,n1,n2)
\newline
     [,1]          [,2]                                       
\newline
[1,] "Unbiased e:" "0.674259756444758"                        
\newline
[2,] "variance:"   "0.41613476136966"                         
\newline
[3,] "CI:"         "[ -0.354146439977423 , 1.65626025590509 ]"
}

This type of functions also has the options for the type I error rate, the biased effect size, and the vector-style output.

\section{Discussion}
\label{discuss}
\subsection{Correspondence of effect sizes and $t$ tests}
Comparison of the effect sizes of the difference and $t$ tests shows the clear correspondence between them (Table 3). Statistic $d$ (5) corresponds to the unpaired two-sample $t$ test (Student, 1908; Fisher, 1925), whose statistic is the basis of $g$ (3). Statistic $e^\text{biased}$ (8) uses the statistic (9) of Welch's $t$ test (Welch, 1947), which aims to test two means with unequal variances, and $c^\text{biased}$ (14) uses the same statistic as the one-sample $t$ test (Fisher. 1925). Considering this, it is natural that power analyses should be conducted, using the corresponding pair of the effect size and $t$ test. In other words, power analyses of Student's one-sample $t$ test, Student's unpaired two-sample $t$ test, and Welch's $t$ test should be conducted based on the $c$ statistic (15), $d$ (5), and the $e$ statistic (11), respectively. Co-use of non-corresponding t test and effect size causes inconsistence of the assumption about the population(s).

\vspace*{0cm}
\begin{table}[h]
\caption{Correspondence of assumptions, t values, and effect sizes of the difference.}
\hspace{-2cm}
\begin{tabular}{c|ccc}
&\shortstack{\large{One sample \&}\\\large{a constant}}&\shortstack{\large{Two samples}\\\large{with equal variance}}& \shortstack{\large{Two samples}\\\large{with unequal variance}} \\ \hline

\rule[-2mm]{0mm}{16mm}\shortstack{\raisebox{1.3em}{\large{Assumption}}}&\shortstack{\raisebox{1.3em}{\large{Normality}}}&\shortstack{\large{Normality,}\\\large{ Independence, \&}\\\large{Equal Variance}}&\shortstack{\raisebox{0.5em}{\large{Normality \&}}\\\ \raisebox{0.5em}{\large{Independence}}} \\

\rule[-6mm]{0mm}{12mm}\shortstack{\large{t value}}&\textit{\large{$t=\frac{\bar{Y^1} - C}{\sqrt{s_1^2/(n_1-1)}}$}}&\textit{\large{$t=\frac{\bar{Y^1} - \bar{Y^2}}{S^\text{pooled}/\sqrt{\tilde{n}}}$}}&\textit{\large{$t=\frac{\bar{Y^1} - \bar{Y^2}}{\sqrt{s_1^2/n_1+s_2^2/n_2}}$}} \\

\rule[-6mm]{0mm}{12mm}\shortstack{\large{Effect size}}&\textit{\large{$c=\frac{\bar{Y^1} - C}{s_1}J(n_1-1)$}}&\textit{\large{$d=\frac{\bar{Y^1} - \bar{Y^2}}{S^\text{pooled}}J(n_1+n_2-2)$}}&\textit{\large{$e=\frac{\bar{Y^1} - \bar{Y^2}}{\sqrt{(s_1^2/n_1+s_2^2/n_2)\tilde{n}}}J(f)$}} \\ 

\end{tabular}

\label{corTES}
\end{table}

\subsection{Effect size and sample size}
Here, the relationship between the effect sizes of the difference and sample sizes is described. The value of $g$ (3), a biased estimator of the effect size of the difference under the equal variance is independent of the sample sizes, when the assumption of the variance equality ($s_1=s_2$) is fulfilled. However, when $s_1\neq s_2$, it depends on the ratio $q=(n_1-1)/(n_2-1)$. This is because $g$ (3) is no longer an estimator of $\delta$ (2) under $s_1\neq s_2$, and it will be an biased estimator of the other parameter $\delta'_q$, which is
\begin{equation*}
\delta'_q=\frac{\mu_1-\mu_2}{\sqrt{(q\sigma_1^2+\sigma_2^2)/(1+q)}}.
\end{equation*}
Note that even $d$ (5) cannot be the unbiased estimator of $\delta'_q$ when $s_1\neq s_2$. This is because $g$ (3) is not distributed as non-central $t$ variate in this situation. Even if $n_1$ and $n_2$ vary, $g$ (3) roughly estimates the same parameter, given the ratio $q$ is fixed.

Next, the $e^\text{biased}$ (8) is an biased estimator of $\epsilon_r$ (7), but $\epsilon_r$ (7) equals to the other parameters in the particular situation. When $s_1=s_2$, $\epsilon_r=\delta$, and $e^\text{biased}$ (8) equals to $g$ (3), and is independent of the sample sizes. When $s_1\neq s_2$ and $n_1=n_2$, $\epsilon_r=\delta'_q$. In this case, $e^\text{biased}$ (8) equals to $g$ (3), and is also independent of the sample sizes. While $d$ (5) is not an unbiased estimator of $\delta'_q$, $e$ (11) is its unbiased estimator. Therefore, usage of $e$ (11) is always preferable to $d$ (5) in this situation. When $s_1\neq s_2$ and $n_1\neq n_2$, $e^\text{biased}$ (8) depends on the rate $r=n_1/n_2$. Therefore, strictly speaking, multiple $e^\text{biased}$s can be comparable only when the sample size ratio $r$ is identical.

Figure 2 shows the behavior of ratio $d/e$ for the different variance $s_2^2$ and sample size $n_1$. When $n_1=n_2$ (Fig. 2. Line 3), this is the same situation as Fig. 2, and $d$ (5) and $e$ (11) are equivalent at $s_1^2=s_2^2$, and $d$ (5) is always larger than $e$ (11) for the other values of $s_2^2$. The ratio $d/e$ reaches at the maximum when $s_2^2 = 0$ and $s_2^2 \to \infty$. The maximum ratio reaches $1.414$ when $n_1=n_2=3$, and the ratio gets smaller for larger $n_1$ and $n_2$. When $n_1>n_2$ (Fig. 2. Line 1 and 2), $d/e$ gets larger for larger $s_2^2$. On the other hand, when $n_1<n_2$ (Fig. 2. Line 4 and 5), $d/e$ gets smaller for larger $s_2^2$.

\begin{figure}[hpt]
\begin{flushleft}
\includegraphics[width=12cm,pagebox=cropbox,clip]{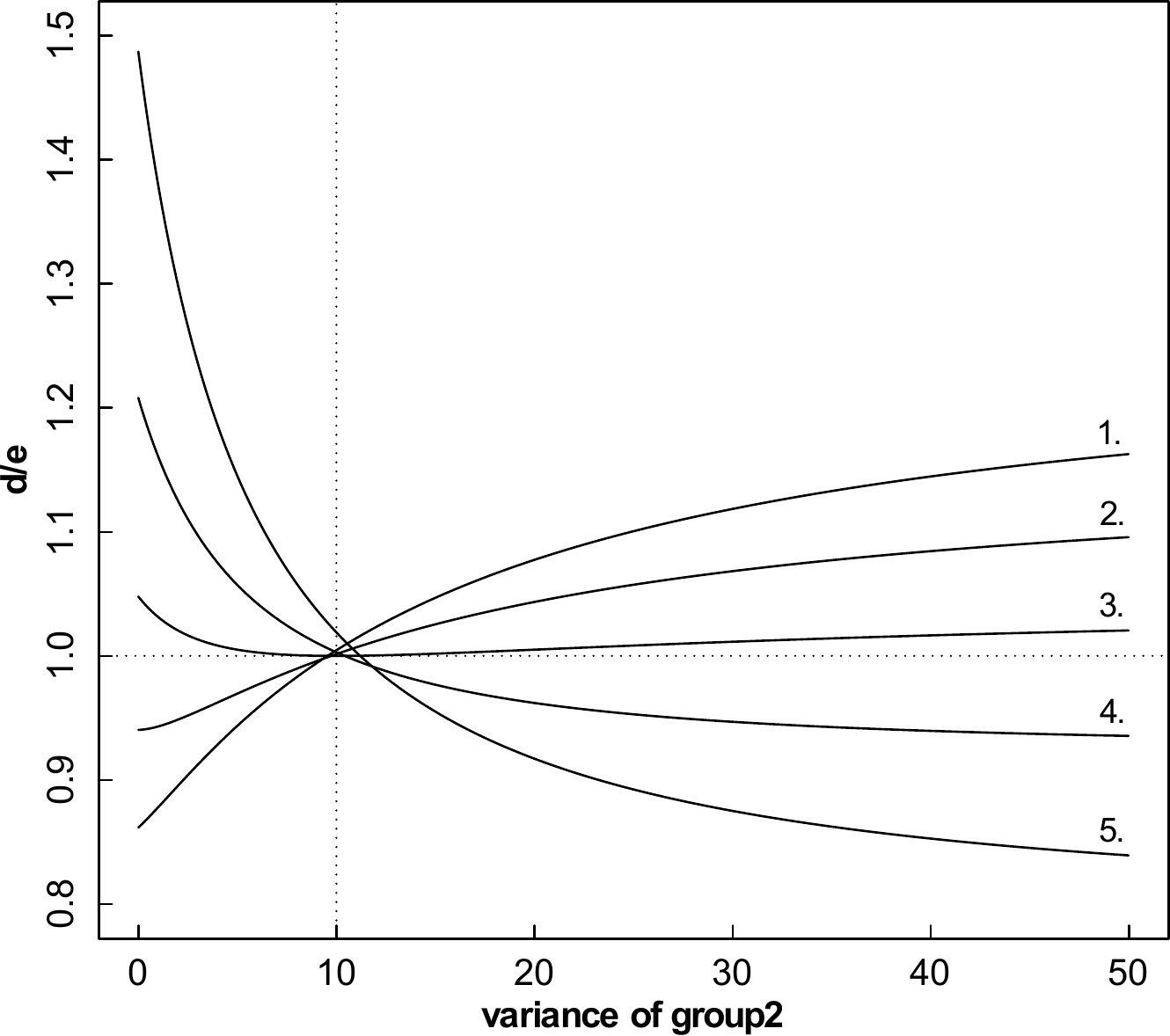}
\end{flushleft}
\caption{Ratio of $d$ (5) to $e$ (11) under $\bar{Y^1}=1$, $\bar{Y^2}=0$, $n_2=10$, $s_1^2=10$, and $s_2^2=0, 0.01, 0.02, ..., 49.99$. Numbered solid lines are $d/e$ for the various $n_1$; Line 1: $n_1=14$. Line 2: $n_1=12$. Line 3: $n_1=10$. Line 4: $n_1=8$. Line 5: $n_1=6$. Dotted lines are $d/e = 1.0$ and $s_2^2=10 (=s_1^2)$.}
\label{DvsE}
\end{figure}

Unlike $g$ (3) or $e^\text{biased}$ (8), $c^\text{biased}$ (14) is always independent of the sample size.

The behavior of the unbiased estimator of the effect sizes ($d$ (5), $e$ (11), and $c$ (15)) are almost identical with those biased, but they slightly increase as the sample sizes get large. This is because of the correction coefficient $J$ (6), and its behavior is illustrated by Hedges (1981) in detail.

\subsection{Potential applications of the new effect sizes}
The effect size $e$ (11) has a vast applicable range covering all kinds of natural and social sciences. This is because $e$ (11) corresponds to Welch's $t$ test, whose use is encouraged over Student's $t$ test these days (e.g. Ruxton (2006)). Especially when the sample sizes of two groups can be fixed, and the variances of them are different, the effect size $e$ (11) is the best suitable option. The effect size $c$ (15) has a relatively narrower range of the application. An effect size concerning constant may be needed in some simulation studies (vs. the optimal value) or physics (vs. physical constant).

\section*{Acknowledgements}

We would like to thank Dr. Fumio Tajima for advising on the English expressions in this article. This study was partly supported by National Bioresource Project from AMED.

\section*{Appendix}

\subsection*{Proofs of unbiasedness and variation of $e$}
In short, this proof is an application of the proof in Hedges (1981) to the statistic $v$ in Welch (1938). Suppose two independently and normally distributed populations $N_1(\mu_1,\sigma_1^2 )$ and $N_2(\mu_2, \sigma_2^2)$. Their sample means are $\bar{Y}^1$ and $\bar{Y}^2$, and their samples are $Y^1_i$ ($i= 1,...,n_1$) and $Y^2_i$ ($i= 1,...,n_2$). The statistic $e^\text{biased}$ (8) between them can be converted into
\begin{equation}
\label{Aone}
\sqrt{\tilde{n}}e^\text{biased}=\frac{(\bar{Y}^1 - \bar{Y}^2)/\sqrt{(\sigma_1^2/n_1)+(\sigma_2^2/n_2)}}{\sqrt{wf/f}},
\end{equation}
where
\begin{equation*}
w=\frac{s_1^2/n_1+s_2^2/n_2}{(\sigma_1^2/n_1)+(\sigma_2^2/n_2)}.
\end{equation*}
Here, since $N_1$ and $N_2$ are independently and normally distributed, the numerator of (16) has the normal distribution of $N(\theta,1)$, where 
\begin{equation*}
\theta=\frac{\mu_1-\mu_2}{\sqrt{(\sigma_1^2/n_1+\sigma_2^2/n_2)}},
\end{equation*}
and the $s_i^2$ is the same as (4).
In the denominator, $wf$ is approximately distributed as $\chi^2 (f)$ (Welch, 1938). Therefore, $\sqrt{\tilde{n}}e^\text{biased}$ is distributed as a non-central $t$ variate with the non-centrality parameter $\theta$ and approximate degree of freedom $f$ (12). From the nature of the non-central $t$ distribution (e.g. Johnson and Welch, 1940), the expected value of $e^\text{biased}$ (8) is
\begin{eqnarray*}
{\rm E}(\sqrt{\tilde{n}}e^\text{biased}) &=&\theta\frac{\sqrt{f/2}\Gamma\{(f - 1)/2\}}{\Gamma(f/2)}\\
{\rm E}(e^\text{biased})&=&\theta/\sqrt{\tilde{n}}/J(f).
\end{eqnarray*}
Now, supposing $r=n_1/n_2$, then $\theta/\sqrt{\tilde{n}}=\epsilon_r$. In this case, the expected value of $e$ (11) is
\begin{eqnarray*}
{\rm E}(e)&=&{\rm E}\{e^\text{biased}J(f)\}\\
&=&{\rm E}(e^\text{biased})J(f)\\
&=&\{\theta/\sqrt{\tilde{n}}/J(f)\}J(f)\\
&=&\epsilon_r.
\end{eqnarray*}
Thus, $e$ (11) is an unbiased estimator of $\epsilon_r$ (7). The variation of $e^{bised}$ (8) is
\begin{eqnarray*}
{\rm var}(\sqrt{\tilde{n}}e^\text{biased})&=&\frac{f}{f-2}(1+\theta^2)-\theta^2/J^2(f)\\
{\rm var}(e^\text{biased})&=&\frac{f}{f-2}(1/\tilde{n}+\theta^2/\tilde{n})-\theta^2/J^2(f)/\tilde{n}.
\end{eqnarray*}
Therefore, the variation of $e$ (11) is
\begin{eqnarray*}
{\rm var}(e)&=&{\rm var}\{e^\text{biased}J(f)\}\\
&=&\frac{f}{f-2}J^2(f)\{1/\tilde{n}+(\theta/\sqrt{\tilde{n}})^2\}-(\theta/\sqrt{\tilde{n}})^2\\
&=&\frac{f}{f-2}J^2(f)(1/\tilde{n}+\epsilon_r^2)-\epsilon_r^2.
\end{eqnarray*}

\begin{flushright}
$\Box$
\end{flushright}
\subsection*{Proofs of unbiasedness and variation of $c$}
The bias correction and derivation of the variance can be proved in the same way as that of $d$ (5). The statistic $c^\text{biased}$ (14) can be converted into
\begin{equation}
\label{A_c}
\sqrt{n-1}c^\text{biased}=\frac{(\bar{Y}^1 - C)}{s/\sqrt{n_1-1}},
\end{equation}
and this (17) is distributed as a non-central $t$ variate with non-centrality parameter
\begin{equation*}
\frac{\mu-C}{\sigma/\sqrt{n_1-1}}
\end{equation*} and degree of freedom $n_1-1$. Therefore, the expected value $c^\text{biased}$ (14) is
\begin{eqnarray*}
{\rm E}(\sqrt{n_1-1}c^\text{biased}) &=&\frac{\mu-C}{\sigma/\sqrt{n_1-1}}\frac{\sqrt{(n_1-1)/2}\Gamma\{(n_1 - 2)/2\}}{\Gamma((n_1-1)/2)}\\
{\rm E}(c^\text{biased})&=&\frac{\mu-C}{\sigma}\frac{1}{J(n_1-1)}\\
{\rm E}(c^\text{biased})&=&\gamma/J(n_1-1).
\end{eqnarray*}
Because $c=c^\text{biased}J(n_1-1)$, the expected value of $c$ (15) is
\begin{eqnarray*}
{\rm E}(c) &=&{\rm E}\{c^\text{biased}J(n_1-1)\}\\
&=&{\rm E}(c^\text{biased})J(n_1-1)\\
&=&\gamma.
\end{eqnarray*}
Thus, $c$ is an unbiased estimator of the effect size parameter $\gamma$ (13). The variation of $c^\text{biased}$ (14) is
\begin{eqnarray*}
{\rm var}(\sqrt{n_1-1}c^\text{biased}) &=&\frac{n_1-1}{n_1-3}\{1+(\frac{\mu-C}{\sigma/\sqrt{n_1-1}})^2\}-(\frac{\mu-C}{\sigma/\sqrt{n_1-1}})^2\frac{1}{J^2(n_1-1)}\\
{\rm var}(c^\text{biased})&=&\frac{n_1-1}{n_1-3}\{\frac{1}{n_1-1}+(\frac{\mu-C}{\sigma})^2\}-(\frac{\mu-C}{\sigma})^2\frac{1}{J^2(n_1-1)}\\
{\rm var}(c^\text{biased})&=&\frac{n_1-1}{n_1-3}(\frac{1}{n_1-1}+\gamma^2)-\gamma^2\frac{1}{J^2(n_1-1)}.
\end{eqnarray*}
Therefore, the variation of $c$ (15) is
\begin{eqnarray*}
{\rm var}(c) &=&{\rm var}\{c^\text{biased}J(n_1-1)\}\\
&=&{\rm var}(c^\text{biased})J^2(n_1-1)\\
&=&\frac{n_1-1}{n_1-3}J^2(n_1-1)(\frac{1}{n_1-1}+\gamma^2)-\gamma^2.
\end{eqnarray*}
\begin{flushright}
$\Box$
\end{flushright}

\subsection*{Proofs of consistency}
First, we treat the proof of $c$ which is simpler than that of $e$. For the proof, we introduce a lemma.
\subsubsection*{Lemma 1}
Assume random samples $Y^1_1, ..., Y^1_n$ from the population with the population mean $\mu_1$ and the population variance $\sigma^2_1$, and consider a parameter $\beta$ and its statistic $b$ = $b(Y^1_1, ..., Y^1_n)$. Then, 

$[b$ is  an unbiased estimator of $\beta$, and $\lim_{n \to \infty}{\rm var}(b)\to 0]$
$\Rightarrow$
$[b$ is a consistent estimator of $\beta$.$]$
\subsubsection*{Proof of lemma 1}
\begin{eqnarray*}
{\rm E}(|b-\beta|^2)&=&{\rm E}(b-\beta)^2+{\rm var}(b-\beta)\\
&=&\{{\rm E}(b)-{\rm E}(\beta) \}^2+{\rm var}(b)\\
&=&\{{\rm E}(b)-\beta \}^2+{\rm var}(b)
\end{eqnarray*}
Given ${\rm E}(b)=\beta$ and $\lim_{n \to \infty}{\rm var}(b)\to 0$,
\begin{equation*}
\lim_{n_ \to \infty}[\{{\rm E}(b)-\beta\}^2+{\rm var}(b)] \to 0.
\end{equation*}
Therefore, $b$ is a mean square consistent estimator of $\beta$, namely, 
\begin{equation*}
\lim_{n_ \to \infty}{\rm E}\{|b-\beta|^2\} \to 0.
\end{equation*}
Here, for an arbitrary positive number $\varepsilon$, by applying Chebyshev's inequality (Chebyshev, 1867), we get
\begin{eqnarray*}
{\rm P}(|b-\beta|\geq\varepsilon)&=& {\rm P}(|b-\beta|^2\geq\varepsilon^2) \\
&\leq&{\rm E}(|b-\beta|^2)/\varepsilon^2
\end{eqnarray*}
From the result shown above, we can say
\begin{equation*}
\lim_{n \to \infty}{\rm E}(|b-\beta|^2)/\varepsilon^2 \to 0.
\end{equation*}
Therefore, using the squeeze theorem, we get
\begin{equation*}
\lim_{n \to \infty}{\rm P}(|b-\beta|\geq\varepsilon) \to 0.
\end{equation*}
Thus, $b$ is a consistent estimator of $\beta$.
$\Box$
\subsubsection*{Proof of consistency of $c$}
Now, let's move on to the proof about $c$ (15). When $n_1 \to \infty$, the variance of $c$ will be
\begin{eqnarray*}
\lim_{n_1 \to \infty}{\rm var}(c)&=& \lim_{n_1 \to \infty}\frac{n_1-1}{n_1-3}J^2(n_1-1)(\frac{1}{n_1-1}+\gamma^2)-\gamma^2 \\
&\to&1\cdot J^2(\infty)(\frac{1}{\infty}+\gamma^2)-\gamma^2 \\
&=&0.
\end{eqnarray*}
Thus, $\lim_{n_1 \to \infty}{\rm var}(c)\to 0$, and $c$ is an unbiased estimator of $\gamma$. Therefore, based on lemma 1, $c$ (15) is a consistent estimator of $\gamma$ (13).
$\Box$

On the other hand, $e$ (11) is consisted of two population. Therefore, a variation of the previous lemma is necessary. 
\subsubsection*{Lemma 2}
Assume two random samples $Y^1_1, ..., Y^1_{n_1}$ and $Y^2_1, ..., Y^2_{n_2}$ from the two mother populations with the population means $\mu_1$ and $\mu_2$, and the population variance $\sigma_1^2$ and $\sigma_2^2$, respectively. Consider a parameter $\beta$ and its statistic $b = b(Y^1_1, ..., Y^1_{n_1}; Y^2_1, ..., Y^2_{n_2})$. Then,

$[b$ is  an unbiased estimator of $\beta$, and $\lim_{(n_1, n_2) \to (\infty, \infty)}{\rm var}(b)\to 0]$
$\Rightarrow$
$[b$ is a consistent estimator of $\beta$.$]$

This lemma can be proved in the same way as lemma 1. 

\subsubsection*{Proof of consistency of $e$}
Now, consider $n_1 = r\phi$ and $n_2=\phi$, to think $\phi \to \infty$, which equals to $(n_1, n_2) \to (\infty, \infty)$. Note that $r>0$ and $\theta>0$, since $n_1 \geq 1$ and $n_2 \geq 1$. Using $r$ and $\phi$, $f$ (6) and $\tilde{n}$ (10) can be expressed as 
\begin{equation*}
f = \frac{(s_1^2/r+s_2^2)^2}{s_1^4/\{r^2(r\phi-1)\}+s_2^4/\{(1/r)^2(\phi-1)\}}
\end{equation*}
and 
\begin{equation*}
\tilde{n} = \frac{r\phi}{r+1}.
\end{equation*}
Therefore, when $\phi \to \infty$, the variance of $e$ (11) will be
\begin{eqnarray*}
\lim_{\phi \to \infty}{\rm var}(e)&=& \lim_{\phi \to \infty}\frac{f}{f-2}J^2(f)(\frac{1}{\tilde{n}}+\epsilon_r^2)-\epsilon_r^2 \\
&=&\lim_{\phi \to \infty}\frac{1}{1-2/f}J^2(f)(\frac{1}{\tilde{n}}+\epsilon_r^2)-\epsilon_r^2 \\
&\to&\frac{1}{1-2/\infty}J^2(\infty)(\frac{1}{\infty}+\epsilon_r^2)-\epsilon_r^2 \\
&=&\frac{1}{1-0}\cdot1\cdot(0+\epsilon_r^2)-\epsilon_r^2 \\
&=&0.
\end{eqnarray*}
The limit does not contain $r$, meaning $\lim_{(n_1, n_2) \to (\infty, \infty)}{\rm var}(e)$ always gives an identical value 0. Also, $e$ is an unbiased estimator of $\epsilon_r$ (7). Therefore, based on lemma 2, $e$ (11) is a consistent estimator of $\epsilon_r$ (7). $\Box$

\label{lastpage}

\begin{thebibliography}{7}
\expandafter\ifx\csname natexlab\endcsname\relax\def\natexlab#1{#1}\fi

\bibitem[\protect\citeauthoryear{Chebyshev}{1867}]{Chebyshev:1867} 
Chebyshev, P. (1867). Des Vaeurs Moyennes. {\it
Journal de Math\'ematiques Pures et Appliqu\'ees} {\bf 12,} 177--184.

\bibitem[\protect\citeauthoryear{Cumming and Finch}{2001}]{Cumming and Finch:2001} 
Cumming, G. and Finch S. (2001). A primer on the understanding, use and calculation of confidence intervals that are based on central and noncentral distributions. {\it
Educational and Psychological Measurement} {\bf 61,} 532--574.

\bibitem[\protect\citeauthoryear{Cohen}{1988}]{Cohen:1988} 
Cohen, J. (1988). {\it Statistical Power Analysis for the Behavioral Sciences}, 2nd edition.
New York: Academic Press.

\bibitem[\protect\citeauthoryear{Fisher}{1925}]{Fisher:1925} 
Fisher, R. A. (1925). {\it Statistical Methods for Research Workers}.
Edinburgh and London: Oliver and Boyd.

\bibitem[\protect\citeauthoryear{Fisher}{1936}]{Fisher:1936} 
Fisher, R. A. (1936). The use of multiple measurements in taxonomic problems.{\it 
Annals of Eugenics}. {\bf 7,} 179--188.

\bibitem[\protect\citeauthoryear{Glass}{1976}]{Glass:1976} 
Glass, G. V. (1976). Primary, secondary, and meta-analysis of research. {\it
Educational Researcher} {\bf 5,} 3--8.

\bibitem[\protect\citeauthoryear{Hedges}{1981}]{Hedges:1981} 
Hedges, L. V. (1981). Distribution theory for Glass's estimator of effect size and related estimators. {\it
Journal of Educational Statistics} {\bf 6,} 107--128.

\bibitem[\protect\citeauthoryear{Hedges and Olkin}{1985}]{Hedges:Olkin:1985} 
Hedges, L. V. and Olkin I. (1985). {\it Statistical Methods for Meta-analysis}.
Orlando: Academic Press.

\bibitem[\protect\citeauthoryear{Johnson and Welch}{1940}]{Johnson:Welch:1940} 
Johnson, N. L. and Welch, B. L. (1940). Applications of the non-central t-distribution. {\it
Biometrika} {\bf 31,} 362--389.

\bibitem[\protect\citeauthoryear{Nakagawa and Cuthill}{2007}]{Nakagawa:Cuthill:2007} 
Nakagawa, S. and Cuthill, I. C. (2007). Effect size, confidence interval and statistical significance: A practical guide for biologists. {\it
Biological Reviews} {\bf 82,} 591--605.

\bibitem[\protect\citeauthoryear{R core team}{2019}]{R:2019} 
R core team. (2019). R: A language and environment for statistical computing. R Foundation for Statistical Computing, Vienna, Austria. URL https://www.R-project.org/.

\bibitem[\protect\citeauthoryear{Ruxton}{2006}]{Ruxton:2006} 
Ruxton, G. D. (2006). The unequal variance t-test is an underused alternative to Student's t-test and the Mann-Whitney U test. {\it
Behavioral Ecology} {\bf 17,} 688--690.

\bibitem[\protect\citeauthoryear{Satterthwaite}{1941}]{Satterthwaite:1941} 
Satterthwaite, F. E. (1941). Synthesis of variance. {\it
Psychometrika} {\bf 6,} 309--316.

\bibitem[\protect\citeauthoryear{Student}{1908}]{Student:1908} 
Student. (1908). The probable error of a mean. {\it
Biometrika} {\bf 6,} 1--25.

\bibitem[\protect\citeauthoryear{Welch}{1938}]{Welch:1938} 
Welch, B. L. (1938). The significance of the difference between two means when the population variances are unequal. {\it
Biometrika} {\bf 29,} 350--362.

\bibitem[\protect\citeauthoryear{Welch}{1947}]{Welch:1947} 
Welch, B. L. (1947). The generalization of `{S}tudent's' problem when several different population variances are involved. {\it
Biometrika} {\bf 34,} 28--35.

\end{thebibliography}
\end{document}